\documentclass[prd,aps,twocolumn,showpacs,floats,floatfix]{revtex4}
\input{epsf}
\usepackage{color,graphicx}
\usepackage{natbib}
\begin{document}
\title{Torsional oscillations of crystalline color-superconducting hybrid
stars: Possible sources for Advanced LIGO? }

\author{Lap-Ming~Lin}
\email{lmlin@phy.cuhk.edu.hk}
\affiliation{Department of Physics and Institute of Theoretical Physics, 
The Chinese University of Hong Kong, Hong Kong, China}

\date{\today}

\begin{abstract}
Deconfined quark matter may exist in a crystalline color-superconducting
phase in the interiors of compact stars. In this paper, we study the 
torsional oscillations of compact stars featuring a crystalline 
color-superconducting quark-matter core in general relativity. 
Depending on the size of the crystalline core 
and the value of the gap parameter $\Delta$,
we find that the frequencies of the torsional oscillation modes
can range from a few hundred hertz to a few kilohertz for our 
canonical $1.4 M_\odot$ compact star models. We have also studied the 
prospect for detecting the gravitational-wave signals emitted from these modes
in a pulsar glitch event. Assuming that at least 10\% of the energy 
released in a Vela glitch can be channeled to the oscillation modes, we find 
that the Einstein Telescope should be able to detect these signals in quite 
general situations. Furthermore, if the size of the crystalline core 
is comparable to the stellar radius and the gap parameter is relatively 
small at $\Delta \sim 5$ MeV, the signal-to-noise ratio for Advanced LIGO 
could reach $\sim$10 for a Vela glitch. 
Our optimistic results suggest that we might already be able to 
probe the nature of crystalline color-superconducting quark matter with the 
second-generation gravitational-wave detectors when they come
online in the next few years. 
\end{abstract}


\maketitle

\section{Introduction}

With their densities reaching a few times the standard nuclear-matter density, 
the dense cores of compact stars have long 
been recognized as the most promising places where different exotic phases
of matter could exist.
In particular, in the early 1970s, it was speculated that the combinations of 
long-range attractive and short-range repulsive channels of neutron-neutron 
interactions might lead to a crystallization of nuclear matter 
(see Chap. 7 of \cite{Haensel_book} for a brief review). 
One of the applications of the possibility of solid cores inside neutron 
stars is the core-quake theory for pulsar glitches 
\cite{Pines72_p83,Baym76_p829}.
It was also pointed out by Dyson already in 1972 that, if solid cores exist
inside neutron stars, core quakes of these stars should excite torsional 
oscillations which might generate detectable gravitational-wave signals on 
Earth \cite{Dyson_remark}.
However, improved nuclear many-body calculations in the middle of 1970s 
essentially ruled out the possibility of crystallization of nuclear matter 
inside neutron stars. 

Another exotic phase of matter inside compact stars proposed in the 1970s is 
the deconfined phase of quark matter 
(e.g., \cite{Ivanenko65,Itoh70,Baym76_p241}). 
Thanks to our improved understanding of the QCD phase diagram nearly forty 
years later today, we now know that the deconfined quark matter inside compact 
stars may also be in a crystalline phase, and hence the possibility of solid 
cores inside compact stars has regained interest in recent years. 
It is now generally believed that the deconfined quarks inside mature 
(cold) compact stars could form Cooper pairs and give rise to 
color superconductivity 
\cite{Alford98_p247,Rapp98_p81,Alford99_p443,Alford03_p074024} 
(see \cite{Alford08_p1455,Anglani13} for reviews). 
At sufficiently high density, pairing between quarks of different colors and
flavors are allowed and the system is said to be in the color-flavor-locked 
(CFL) phase \cite{Alford99_p443}. 
However, at intermediate densities relevant to compact stars, 
the more favored phase is the crystalline color-superconducting quark matter 
with broken spatial symmetries \cite{Alford01_p074016,Casalbuoni05_p89,
Mannarelli06_p114012,Rajagopal06_p094019,Casalbuoni06_p350}. 
The presence of this crystalline phase of quark matter in the core of compact 
stars should produce astrophysical signatures that are very different from 
those of traditional neutron stars with a fluid core. Being able to identify 
these signatures from observational data would provide us a unique 
way to probe the nature of QCD in the high, but not asymptotically high, 
density regime. In this work, we shall focus on the gravitational-wave 
signatures of these objects. 

It has long been known that the gravitational-wave signals emitted from 
compact stars carry important information about the internal structure of the 
stars (see, e.g., \cite{Andersson96_p4134,Kokkotas01_p307,Benhar04_p124015,
Tsui05,Lau10}), 
and thus their detection would give us viable information 
on dense matter. Regarding compact stars featuring a core of crystalline 
color-superconducting quark matter, the gravitational-wave emission due to 
nonaxisymmetric distortions of the solid core have been studied 
and analyzed with the (nondetection) results of the LIGO scientific runs 
S3/S4 \cite{Lin07,Haskell07_p231101} and S5 \cite{Knippel09_p083007}. 
A general conclusion obtained from these works is that the gravitational-wave 
signals are within the reach of LIGO if the solid core is maximally strained. 
In this paper, on the other hand, we shall study the gravitational-wave 
signals emitted from the torsional oscillations of the solid core. 
As we shall see below, depending on the gap parameter and the size of the 
core, the frequencies of these oscillation modes can range from a few hundred 
hertz to a few kilohertz. 
This frequency range is quite different from those standard fluid modes 
of traditional neutron stars, such as the $f$ and $p$ modes, with typical
frequencies at a few kilohertz \cite{Kokkotas99_p2}. 
Their relatively low frequencies also put the torsional oscillation modes 
within the best sensitivity region of ground-based gravitational-wave
detectors such as LIGO, VIRGO, and KAGRA.

The plan of the paper is as follows. In Sec.~\ref{sec:formulation}, we 
summarize the formulation that we employ to calculate the torsional 
oscillations of compact stars. Section~\ref{sec:micro_input} discusses the 
microphysical input that we need for constructing background stellar models 
and mode calculations. We present our numerical results in 
Sec.~\ref{sec:num_results} and discuss the prospect for detecting 
the gravitational-wave signals in Sec.~\ref{sec:GW}. Finally, we conclude
our paper in Sec.~\ref{sec:conclude}. Unless otherwise noted, we use geometric
units with $G=c=1$.

\section{Formulation}  
\label{sec:formulation}

The formulation for studying torsional oscillations of compact stars in 
general relativity was first developed by Schumaker and Thorne 
\cite{Schumaker83} thirty years ago. 
A more recent gauge-invariant formulation of the problem can be found in 
\cite{Karlovini07_p3171}. 
Due to their relevance to the quasiperiodic oscillations
in giant flares emitted from soft gamma-ray repeaters
\cite{Israel05_L53,Watt07_p1446}, torsional oscillations in the solid crust of 
neutron stars have been studied in great detail recently 
(e.g., \cite{Samuelsson07_p256,Sotani07_p261,Vavoulidis07,Sotani08_L5,
Colaiuda11_p3014,Sotani12_p201101,Gabler12_p2054}). 
As far as we are aware, all these works are done within the 
Cowling approximation where the metric perturbations are neglected. 
This should be a good approximation in the study of oscillations in the 
crust of neutron stars, since the density is low in the crust and the 
spacetime variations should be small there. However, this is in general 
not necessarily true for the oscillations of a massive solid core. 
Hence, we shall employ the fully relativistic formulation of 
Schumaker and Thorne~\cite{Schumaker83} to study the torsional oscillations, 
though we shall also compare the relativistic results with those obtained by 
the Cowling approximation. 
  
In the following, we shall first summarize the standard set of equations for 
constructing the unperturbed equilibrium static configuration. 
We then briefly discuss the set of perturbation equations and numerical 
scheme for solving the torsional oscillation modes. 
We refer the reader to the original work \cite{Schumaker83} for the full
derivations.

\subsection{Equilibrium static background}

The unperturbed background is assumed to be a static and spherically symmetric
spacetime described by the metric 
\begin{equation}
ds^2 = - e^{2 \Phi(r)} dt^2 + e^{2 \Lambda(r)} dr^2 + r^2 (d\theta^2 
+ \sin^2 \theta d \phi^2 ) , 
\label{eq:backgd_metric}
\end{equation}
where the functions $\Phi(r)$ and $\Lambda(r)$ depend only on the radial 
coordinate $r$. The equilibrium structure of a compact star is determined
by the standard Tolman-Oppenheimer-Volkov equations: 
\begin{eqnarray} 
{dm \over dr} &=& 4 \pi r^2 \rho , \\
&& \cr  
{dP \over dr} &=& - { (\rho + P) ( m + 4 \pi r^3 P) \over 
r^2 ( 1 - 2 m/ r) } , \\
&&  \cr
{d \Phi \over dr} &=& { m + 4 \pi r^3 P \over r^2 ( 1 - 2m/r) } , 
\end{eqnarray}
where $\rho$ and $P$ are the energy density and pressure of the fluid, 
respectively. The function $m(r)$ is defined by 
$e^{-2\Lambda(r)} = 1 - 2 m(r) / r$. 
With a given equation of state (EOS) $P(\rho)$, the above system 
of differential equations can be solved by imposing the boundary conditions 
(i) $m(0)=0$; (ii) $P(R)=0$; (iii) $e^{2 \Phi(R)} = 1 - 2 M /R$, where
$M=m(R)$ is the total mass of the star and $R$ is its radius.

\subsection{Torsional perturbations}
\label{sec:perturb_eq}

In studying the oscillation modes of compact stars, the metric and fluid 
perturbations are decomposed onto the basis of spherical harmonics
[with indices $(l,m)$].
Without loss of generality, the study can be 
restricted to the $m=0$ modes because the unperturbed background is 
spherically symmetric. Furthermore, the perturbations are in general 
distinguished into two classes according to their parities
\cite{Thorne67_p591}: the so-called polar modes [with $(-1)^l$ parity] 
and axial modes [with $(-1)^{l+1}$ parity]. 
The torsional oscillation modes are of the latter class and hence we shall 
discuss the axial perturbations only. In the so-called Regge-Wheeler 
gauge, the axial metric perturbations $h_{\alpha\beta}$ can be expressed as
\begin{widetext}
\begin{equation}
 h_{\alpha\beta} = \left[\matrix{0&0&0&
            -r^2 {\dot y}(t,r) \cr 
            0&0&0& -r e^{\Lambda-\Phi} Q(t,r) \cr 
            0&0&0&0\cr -r^2 {\dot y}(t,r) 
            & - r e^{\Lambda-\Phi} Q(t,r) &0&0 } \right] 
  \sin \theta {\partial \over \partial \theta} P_l(\cos \theta) , 
\end{equation}
\end{widetext}
where $P_l(\cos \theta)$ is the Legendre polynomials of order $l$ and 
the dot over $y(t,r)$ refers to derivative with respect to $t$. 
The functions $\Lambda$ and $\Phi$ are the background metric functions 
defined in Eq.~(\ref{eq:backgd_metric}). On the other hand, the axial 
fluid perturbation is characterized by the displacement vector
\begin{equation} 
\xi^r = 0 , \ \ \xi^\theta = 0 , \ \ 
\xi^\phi = {Y(t,r) \over \sin \theta } {\partial \over \partial\theta}
P_l(\cos \theta) . 
\end{equation}
To linear order in $\xi^i$, the four-velocity of the fluid is given by
$u^\alpha = e^{-\Phi}( 1, 0, 0, {\dot \xi}^\phi )$.

The stress-energy tensor $T_{\alpha\beta}$ of the matter inside the star 
in general contains two contributions: 
$T_{\alpha\beta} = T_{\alpha\beta}^{\rm bulk} + T_{\alpha\beta}^{\rm shear}$.
The bulk part $T_{\alpha\beta}^{\rm bulk}$ is assumed to take the perfect 
fluid form 
\begin{equation}
T_{\alpha\beta}^{\rm bulk} = (\rho + P) u_{\alpha} u_{\beta} + 
P g_{\alpha\beta} . 
\end{equation}
The shear part $T_{\alpha\beta}^{\rm shear}$ is given by 
\begin{equation}
T_{\alpha\beta}^{\rm shear} = - 2 \mu S_{\alpha\beta} , 
\end{equation}
where $\mu$ is the shear modulus and $S_{\alpha\beta}$ is the shear tensor
which describes the deformations generated by the displacement $\xi^i$. 
The explicit expressions for $S_{\alpha\beta}$ can be found in 
\cite{Schumaker83}. 
In this work, we will consider star models in which there exists two 
regions: a crystalline quark-matter core and a nuclear-matter fluid envelope. 
The shear modulus $\mu$, and hence $T_{\alpha\beta}^{\rm shear}$, is nonzero 
only in the crystalline core. 
It should also be noted that $\rho$ and $P$ cannot support axial perturbations
because they are scalar fields. Thus, our compact star models can support 
torsional oscillations only in their crystalline cores. Of course, a more 
realistic compact star model may also have a thin crust near the surface
to support torsional oscillations. 
Since our focus is on the emitted gravitational-wave signals associated with 
torsional oscillations, which are dominated by the massive crystalline 
core, we shall thus neglect the crust in our study.

Assuming a time dependence $e^{i\omega t}$ for the fluid and metric 
perturbations [e.g., $Y(t,r) = Y(r)e^{i\omega t}$], the linearized 
fluid and Einstein field equations reduce to the following differential 
equations:
\begin{eqnarray}
X^\prime &=& r^4 e^{\Phi+\Lambda} \left[ 16\pi (\rho+P) + 
{(l+2)(l-1)\over r^2} \right. \cr 
&&\cr
&&\left. - { {\rho + P} \over \mu } \omega^2 e^{-2\Phi} \right] \mu Y 
+ ( \mu - \rho - P) r^3 e^{\Lambda-\Phi} Z   \cr
&&\cr
&& + \left[ r \mu^\prime + 3 \mu - (\rho + P) \right] r^2 Q  , 
\label{eq:X_prime} \\
&&\cr
Z^\prime &=&  16\pi r \left( \mu e^{2\Phi}\right)^\prime Y
+ e^{\Phi+\Lambda} \left[ 16\pi \mu + {l(l+1) \over r^2} \right. \cr
&&\cr
&&\left. + 4\pi (\rho - P) - {6 m \over r^3} 
- \omega^2 e^{-2\Phi} \right] Q   , 
\label{eq:Z_prime}
\end{eqnarray}
where primes denote derivatives with respect to $r$. 
The auxiliary functions $X$ and $Z$ are defined by 
\begin{eqnarray}
X &=& \mu r^4 e^{\Phi-\Lambda} Y^\prime , 
\label{eq:X_def} \\
&&\cr
Z &=& e^{\Phi-\Lambda} Q^\prime . 
\label{eq:Z_def} 
\end{eqnarray} 
The above first-order system of equations are equivalent to 
Eqs. (50a) and (50b) in \cite{Schumaker83}, though being expressed in terms of 
different variables. 
After solving the above equations for the mode frequency $\omega$ 
with appropriate boundary conditions (see below), the remaining 
metric perturbation function $y$ can then be obtained by 
\begin{equation}
y = {1\over \omega^2} \left[ { - e^{\Phi-\Lambda} \over r^2 }
( r Q)^\prime + 16 \pi \mu e^{2\Phi} Y \right] . 
\label{eq:y_IVP}
\end{equation}
The fact that $y$ is decoupled from the wave equations 
(\ref{eq:X_prime}) and (\ref{eq:Z_prime}), and being determined separately by 
Eq.~(\ref{eq:y_IVP}), is traceable to the gauge choice used in the 
formulation as discussed in \cite{Karlovini07_p3171}. 
As we are only interested in obtaining the mode frequency $\omega$, we shall 
not discuss the function $y$ any further.

In the nuclear-matter fluid region outside the core, the shear modulus 
vanishes and the fluid cannot support torsional oscillations. 
In this region, the fluid only differentially rotates and the gravitational 
perturbation $Q$ is completely decoupled from the fluid \cite{Thorne67_p591}. 
It can be shown that the governing equation for $Q$ [Eq.~(\ref{eq:Z_prime})]
in this region can be written as (with $\mu=0$)
\begin{equation}
{d^2Q\over dr_*^2} = {e^{2\Phi}\over r^3} \left[ l(l+1) r + 
4\pi r^3 (\rho - P) - 6 m \right]Q - \omega^2 Q , 
\label{eq:RW_eq_inside}
\end{equation} 
where the Regge-Wheeler radial coordinate $r_*$ is defined by 
$\partial /\partial r_* = e^{\Phi - \Lambda} {\partial/ \partial r}$.
This is the standard wave equation governing the so-called axial $w$ modes
of neutron stars, which are due to spacetime oscillations
(see, e.g., \cite{Kokkotas99_p2} for a review).
Outside the star where $\rho$ and $P$ vanish, Eq.~(\ref{eq:RW_eq_inside})
reduces to the Regge-Wheeler equation governing the perturbations of 
Schwarzschild spacetime:
\begin{equation}
{ {d^2 Q} \over {d r_*^2} } 
= \left( 1 - {2 M \over r} \right)
\left[ { l(l+1)\over r^2 } - { 6 M\over r^3 } \right] Q - \omega^2 Q , 
\label{eq:Regge_Wheeler}
\end{equation}
where $r_* = r + 2M \ln (r/2M - 1)$ and $M=m(R)$ is the mass of the star.

\subsection{Boundary conditions and numerical methods}
\label{sec:BC}

In order to calculate the torsional oscillation modes, the perturbation 
equations as listed in Sec.~\ref{sec:perturb_eq} must be solved with suitable 
boundary conditions at the center, the solid-fluid interface, the 
stellar surface and at infinity. 
First, the regularity conditions of the fields at the 
center require that 
\begin{equation} 
Y \sim r^{l-1} , \ \ Q \sim r^{l+1} , 
\end{equation} 
near $r=0$. At the solid-fluid interface $r=R_c$, we impose the zero-traction
condition which is given by \cite{Schumaker83}
\begin{equation}
\mu \left( Y^\prime - e^{\Lambda-\Phi} Q / r \right) = 0 .
\label{eq:interface_bc}
\end{equation} 
On the other hand, the continuity conditions of the intrinsic and extrinsic 
curvatures require $Q$ to be continuous there and also at the stellar surface
$r=R$. Outside the star, the Regge-Wheeler equation (\ref{eq:Regge_Wheeler})
has the asymptotic solution 
\begin{equation} 
Q \sim A_{\rm in} e^{i\omega r_*} + A_{\rm out} e^{-i\omega r_*} ,
\label{eq:Q_infty}
\end{equation}
as $r_* \rightarrow \infty$. The quasinormal mode frequency $\omega$ is 
determined by requiring that there is only outgoing gravitational radiation at 
infinity (i.e., $A_{\rm in}=0$). 
Note that $\omega$ is in general a complex number because of the damping 
due to the emission of gravitational waves. 

In the following, we outline our numerical scheme to obtain the 
mode frequency. Given an equilibrium background configuration and a trial 
frequency $\omega$, we integrate Eqs.~(\ref{eq:X_prime})-(\ref{eq:Z_def})
in the solid core using initial conditions $Y(r)=Y_0 r^{l-1}$ and 
$Q(r)=Q_0 r^{l+1}$, where the ratio $Y_0/Q_0$ is set to some value. 
The integration is carried out to the solid-fluid interface $r=R_c$ at which 
we check whether the boundary condition there [Eq.~(\ref{eq:interface_bc})] 
is satisfied. If Eq.~(\ref{eq:interface_bc}) is not satisfied, a new value 
of $Y_0/Q_0$ is used and the integration is repeated. Once a correct value
of $Y_0/Q_0$ is found, we then integrate Eq.~(\ref{eq:RW_eq_inside}) in the 
fluid envelope and finally Eq.~(\ref{eq:Regge_Wheeler}) in the vacuum. 
The trial frequency $\omega$ is the desired quasinormal mode solution if 
the asymptotic ingoing-wave amplitude $A_{\rm in}$ vanishes. In practice, we 
determine $A_{\rm in}$ from Eq.~(\ref{eq:Q_infty}) at $r=100R$ and the real 
parts of the mode frequencies are determined by the locations of 
deep minima in a graph of $\log_{10} |A_{\rm in} |$ vs ${\rm Re}(\omega)$.  

\begin{figure}
\centering
\includegraphics*[width=8cm]{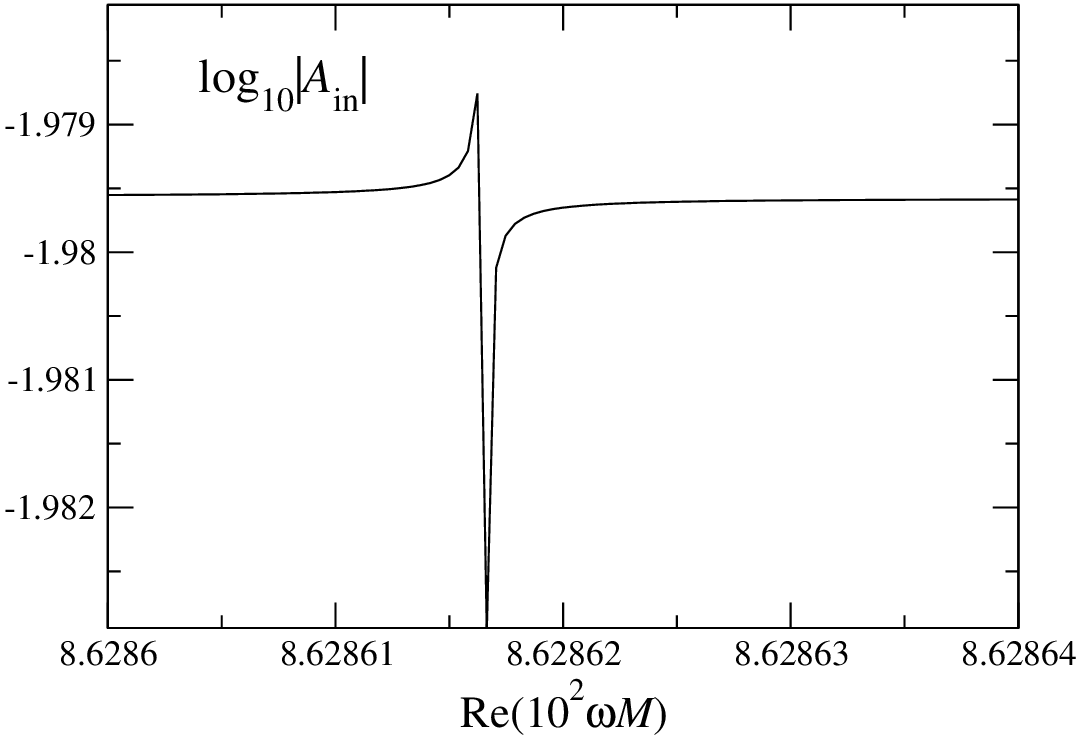}
\caption{Plot of $A_{\rm in}$ against ${\rm Re}(\omega M)$ for a typical 
oscillation-mode calculation in this work. The location of the minimum gives
the frequency of a mode. }
\label{fig:modelA_gap10_3rdmode}
\end{figure}

As a demonstration of the graphical method, we show a typical result in our 
calculations in Fig.~\ref{fig:modelA_gap10_3rdmode}. 
The location of the minimum in the figure corresponds to the frequency of a 
torsional oscillation mode of the stellar model A (discussed in 
Sec.~\ref{sec:num_results}) with the gap parameter $\Delta=10$ MeV. 
We refer the reader to \cite{Comer99,Andersson02_p104002,Lin08} for similar 
applications of this graphical method. 
Since the torsional oscillation modes emit current-quadrupole gravitational 
waves, these modes are damped much more slower than 
polar fluid modes such as the $f$ and $p$ modes (as they emit
mass-quadrupole waves). 
Hence, we need to use a high resolution along the $\omega$ axis 
in order to locate the modes. We typically set $\Delta \omega M < 10^{-7}$
in the calculations, where $\Delta \omega$ is the step size along the 
$\omega$ axis.

\section{Microphysics input}
\label{sec:micro_input}

\subsection{Equation of state}
\label{sec:eos}

In order to construct equilibrium hybrid stars, we need to provide 
EOS models to describe the quark-matter core and nuclear-matter envelope. 
The quark-matter EOS is in general derived using the MIT bag model and its
variations or more realistic Nambu-Jona-Lasinio (NJL) type models
\cite{Buballa05_p205}. 
In our numerical calculations we use the simple phenomenological quark-matter 
EOS model of Alford {\it et al.} \cite{Alford05_p969} for the quark core. 
The model is based on the thermodynamic potential
\begin{equation}
\Omega_{\rm QM} = - { 3 \over 4 \pi^2} a_4 \mu_q^4 
+ {3 \over 4 \pi^2} a_2 \mu_q^2 + B_{\rm eff} , 
\label{eq:quark_eos}
\end{equation}
where $\mu_q$ is the quark chemical potential. The phenomenological 
parameters $a_4$, $a_2$, and $B_{\rm eff}$ are independent of $\mu_q$.
The parameter $a_4 (\le 1)$ is used to account for nonperturbative
QCD corrections. The limiting case $a_4=1$ corresponds to 
three flavors of noninteracting quarks. The reasonable value for $a_4$ is 
expected to be of order 0.7 \cite{Alford05_p969}. 
The parameter $a_2$ is used to model the effects of quark masses and pairing. 
Finally, the effective bag constant $B_{\rm eff}$ can be regarded as a 
parameter to control the density at which the transition from 
nuclear matter to quark matter occurs. 

In the nuclear-matter envelope we use the model of Akmal {\it et al.} 
(APR) \cite{APR} at high densities. At lower densities, 
the APR EOS is matched to the model of Douchin and Haensel \cite{SLY4}, 
which is followed by the model of Baym {\it et al.} \cite{BPS} and 
Haensel and Pichon \cite{Haensel94_p313}. 
We implement the phase transition from nuclear matter to deconfined
quark matter using a Maxwell construction. The EOS for the two different 
phases are matched by requiring that the pressures of both phases are equal at 
certain (baryonic) chemical potential.

We could in principle employ more sophisticated models for the quark core, 
such as the NJL model used in \cite{Ippolito08_p023004,Knippel09_p083007} 
where the three-flavor crystalline phase of QCD is built in consistently. 
At this point we are more concerned with the technical developments 
of the numerical program and the qualitative properties of the torsional 
oscillation modes, so we focus only on the phenomenological model 
[Eq.~(\ref{eq:quark_eos})] as an illustrative example in this work.

\subsection{Shear modulus} 
\label{sec:shear_modulus} 

The shear modulus of crystalline color-superconducting quark matter has been
calculated by Mannarelli {\it et al.} \cite{Mannarelli07_p074026} and is 
given by 
\begin{equation} 
\mu = 2.47 {\rm \ MeV/fm^3} \left( {\Delta\over 10{\ \rm MeV} } \right)^2
\left( {\mu_q \over 400 {\ \rm MeV} }\right)^2 ,
\label{eq:mu}
\end{equation}
where $\Delta$ is the gap parameter. 
It should be pointed out that the result is obtained by 
performing a Ginzburg-Landau expansion to order $\Delta^2$. 
Since the control parameter for the expansion is about $1/2$ 
\cite{Mannarelli07_p074026}, Eq.~(\ref{eq:mu}) can thus be regarded as an 
estimation of $\mu$ only. 
For quark matter inside compact stars, the quark chemical potential $\mu_q$ is 
expected to lie in the range $350\ {\rm MeV} < \mu_q < 500\ {\rm MeV}$
\cite{Mannarelli07_p074026,Ippolito07_p036}. 
The gap parameter $\Delta$ is less certain and is expected to lie between 
10 and 100 MeV in the CFL phase where the strange quark mass can be neglected. 
However, Mannarelli {\it et al.} \cite{Mannarelli07_p074026} 
estimate that $\Delta$ should be in the range $ 5\ {\rm MeV} \lesssim \Delta
\lesssim 25\ {\rm MeV}$ in order for the quark matter to be in the crystalline
phase rather than the CFL phase. 
As a result, the shear modulus of crystalline color-superconducting 
quark matter is in the range $0.47\ {\rm MeV/fm}^3 < \mu < 
24\ {\rm MeV/fm}^3$, which is much larger than that of the neutron star's 
crust [$\sim O(1\ {\rm keV/fm}^3)$]. This is the reason why the maximum 
equatorial ellipticity sustainable by stellar models with a crystalline 
color-superconducting quark-matter core can be a few orders of magnitude 
larger than which can be supported by the neutron star's 
solid crust \cite{Lin07,Haskell07_p231101,Knippel09_p083007} 
(see also \cite{Owen05,Johnson13_p044004} for relevant studies).

\begin{table}
 \begin{tabular}[t]{|c|c|c|c|c|c|c|}
\hline
Model & $a_2^{1/2}$  & $a_4 $ & 
$B_{\rm eff}^{1/4}$  & $\rho_c $ & $R_c$  & $R$   \\ 
      &  (MeV)  &    &  (MeV) &  
$(10^{15} {\rm g\ cm^{-3} })$ &  (km) &  (km)  \\ 
\hline
A  & 100  & 0.85  & 160  & 1.487  & 8.28   & 10.31   \\
\hline
B  & 100  & 0.8  &  160  & 1.265  & 4.71   & 11.07   \\ 
\hline 
C  & 200  & 0.9  &  150  & 1.205  & 1.71    & 11.24   \\ 
 \hline
 \end{tabular}
 \caption{The canonical background stellar models for the oscillation-mode
calculations as discussed in the text. 
All models have the same total mass $1.4 M_\odot$.  }
 \label{tab:models}
\end{table}

\section{Numerical Results}
\label{sec:num_results}

\begin{figure}
\centering
\includegraphics*[width=8cm]{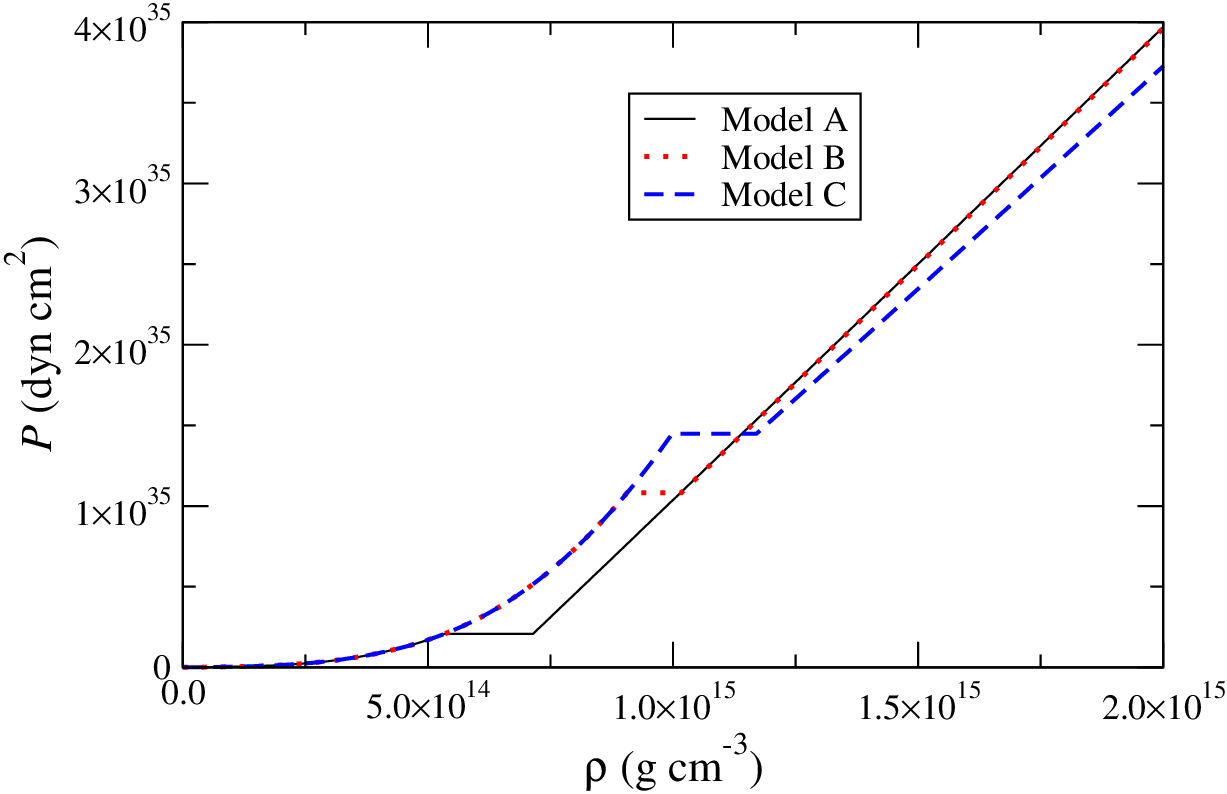}
\caption{Pressure is plotted against energy density for our EOS models.}
\label{fig:P_rho}
\end{figure}

\begin{figure}
\centering
\includegraphics*[width=8cm]{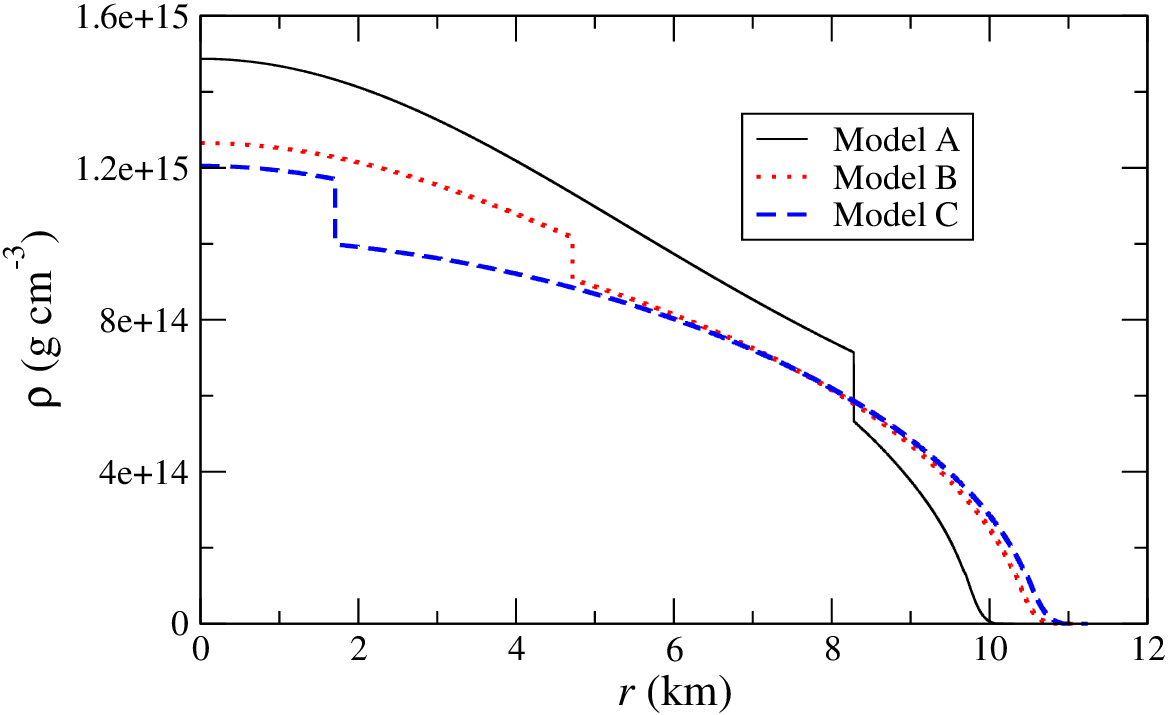}
\caption{Plot of the density profiles for the three background stellar models 
listed in Table~\ref{tab:models}. }
\label{fig:rho_profile}
\end{figure}

\subsection{Equilibrium stellar models} 
\label{sec:models}

In the oscillation-mode calculations, we need to first construct 
equilibrium stellar models which depend on the background EOS model. 
While the EOS for the nuclear fluid envelope is fixed in our calculations, 
there are still three free parameters ($a_2$, $a_4$, $B_{\rm eff}$) in the 
quark-matter EOS for the core. Since the parameter space is vast and we are 
mainly interested in understanding the qualitative properties, we shall only 
use three different sets of parameters to construct three ``canonical'' 
background stellar models with the same total mass $M=1.4 M_\odot$.  
The EOS parameters ($a_2$, $a_4$, $B_{\rm eff}$), central density 
$\rho_c$, quark-core radius $R_c$, and stellar radius $R$ of the background 
models are summarized in Table~\ref{tab:models}. 

The EOS parameters are chosen in order to produce quark-matter EOS models 
which can match to our fixed nuclear-matter EOS at low density as discussed
in Sec.~\ref{sec:eos}. 
Furthermore, the three sets of parameters are also chosen to yield significant 
differences in the internal structures of the stellar models for comparison. 
The effects of different parameters on the EOS can be seen in 
Fig.~\ref{fig:P_rho}. 
In the figure, we plot the pressure against energy density for the three
EOS models. It is note that, according to the Maxwell construction of the 
phase transition, there is in general a discontinuity in the energy density at 
constant pressure as shown in Fig.~\ref{fig:P_rho} 
(see also \cite{Alford05_p969}).  
Figure~\ref{fig:P_rho} shows that the density at which the transition from 
nuclear matter to quark matter occurs, and also the jump in the density at the 
transition, depend quite sensitively on the EOS parameters. 
This leads to significant differences in the internal structures of 
the three background stellar models. 
In particular, by comparing models A and B, it can been seen that a small 
change in the value of $a_4$ can lead to a large difference in the density 
at which the phase transition occurs. 
In Fig.~\ref{fig:rho_profile} we plot the density profiles of the three 
background models for comparison. In our oscillation-mode calculations, we 
shall consider how the size of the quark core for a $1.4 M_\odot$ 
canonical stellar model affects the mode frequency. 
Model A thus represents the situation of a large quark core 
($R_c/R\sim 0.8$), while model C represents the more conservative case of 
a small core ($R_c/R \sim 0.1$).

\begin{figure}
\centering
\includegraphics*[width=8cm]{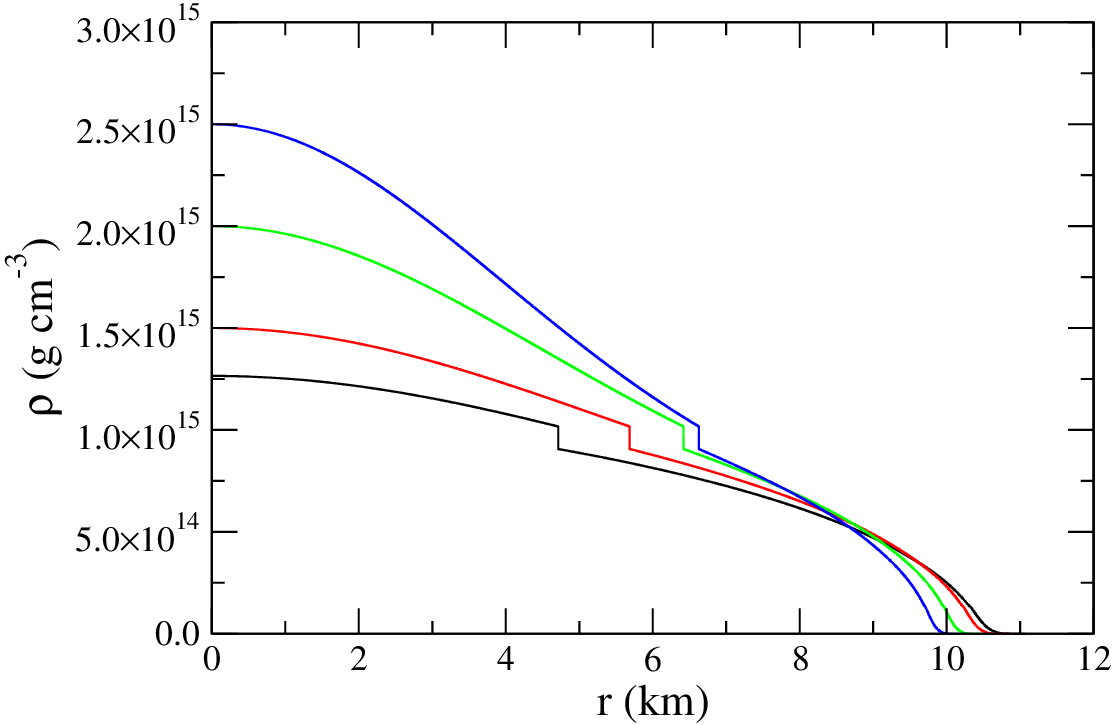}
\caption{ Density profiles for stellar models with
different central densities. The stars are constructed with the same 
EOS model B.  
The star with the lowest central density corresponds 
to our canonical model B listed in Table~\ref{tab:models}. 
The star with the highest central density has a mass $1.63M_\odot$. }
\label{fig:rho_eosB_diffM}
\end{figure}

While we shall only focus on the three $1.4M_\odot$ stellar models in the
oscillation-mode calculations, it is also interesting to see how the stellar
structure would change for more massive configurations. 
In Fig.~\ref{fig:rho_eosB_diffM}, we plot the density profiles for stellar 
models with different central densities. The stars are constructed with the
same EOS model B. 
The profile with the largest central density in the figure corresponds to 
a $1.63 M_\odot$ hybrid star.
We note that for larger central densities, and hence more massive stars, 
the radius of the quark core becomes less sensitive to the central density
and saturates at $R_c \sim 6.6$ km. We refer the reader to 
\cite{Knippel09_p083007} for a similar trend observed in a NJL model 
where $R_c$ saturates at 7 km for massive stars near $2 M_\odot$.

Although our quark-matter EOS is based only on a phenomenological model, we 
believe that the resulting three stellar models are quite generic in
representing the internal structures of crystalline color-superconducting
hybrid stars. 
Our models are also compatible to the results obtained by the NJL 
model of \cite{Ippolito08_p023004,Knippel09_p083007} in which it is shown 
that large crystalline quark cores ($R_c/R > 0.5$) can exist in compact 
stars.

\begin{figure}
\centering
\includegraphics*[width=8cm]{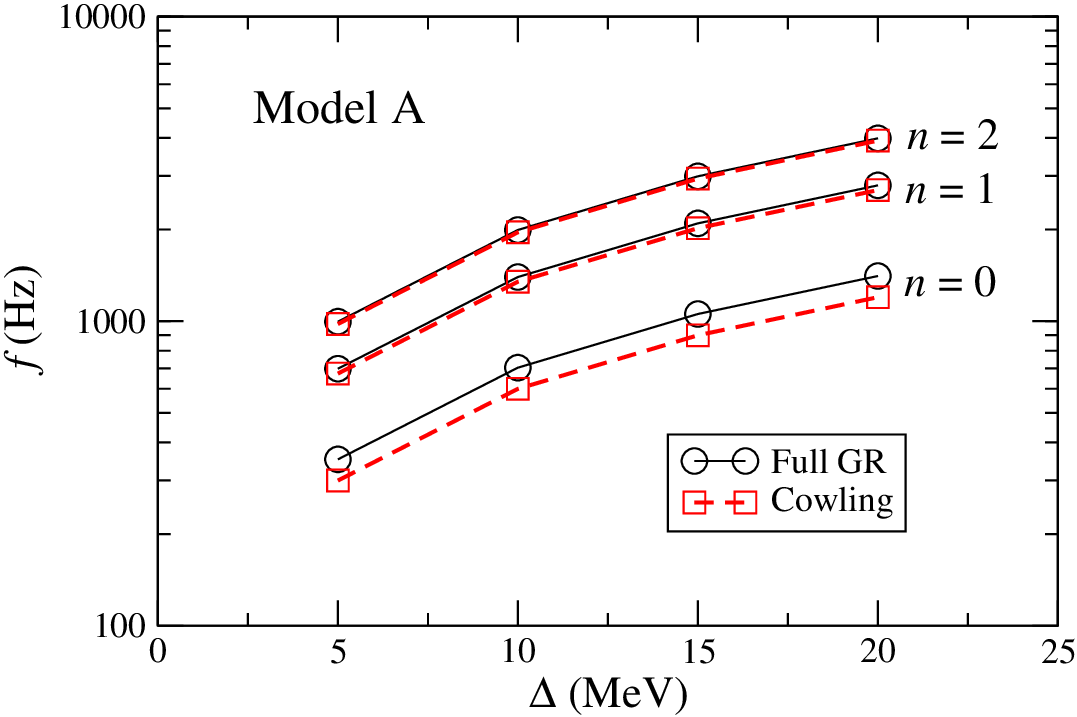}
\caption{ Frequency of the torsional oscillation modes as a function of the 
gap parameter for the stellar model A listed in Table~\ref{tab:models}. The 
integer $n$ is the mode order (i.e., $n=0$ corresponds to the fundamental 
mode and so on). The solid lines are computed using the fully relativistic 
framework as discussed in the text, while the dashed lines are obtained 
by neglecting the metric perturbations (Cowling approximation).  }
\label{fig:modelA_fvsDelta}
\end{figure}

\subsection{Torsional oscillation modes} 
\label{sec:torsion_modes}

In this subsection we consider the $l=2$ torsional oscillation modes of the 
three canonical stellar models presented in Sec.~\ref{sec:models}. 
Our focus will be on how the size of the quark core and the gap parameter 
$\Delta$ affect the frequencies of the modes. In particular, we only consider 
the range $5\ {\rm MeV} \lesssim \Delta \lesssim 25\ {\rm MeV}$ in our 
calculations as this is the theoretically allowed range as discussed in 
Sec.~\ref{sec:shear_modulus}.

\begin{figure}
\centering
\includegraphics*[width=7cm]{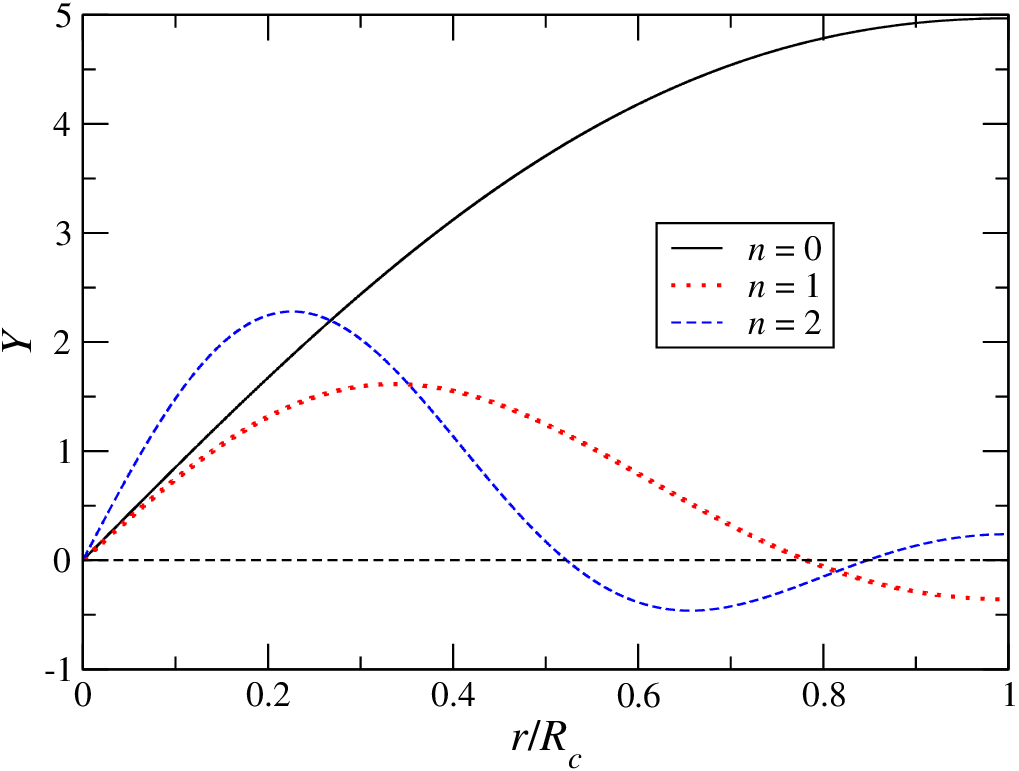}
\caption{ 
Radial amplitudes of the fluid perturbation function $Y$ inside the 
quark core for the torsional oscillation modes of stellar model A and 
$\Delta=10$ MeV. 
The horizontal dashed line is used to pinpoint where the 
amplitudes vanish.}
\label{fig:Yfcn_modelA_gap10}
\end{figure}

We begin by plotting in Fig.~\ref{fig:modelA_fvsDelta} the frequencies of 
the first three torsional oscillation modes of stellar model A against 
$\Delta$. In the figure, the fundamental modes are given by the curves labeled 
as $n=0$. 
Similarly, the first and second harmonics are given by the curves
labeled as $n=1$ and $n=2$, respectively. 
The general pattern of the mode eigenfunctions can be seen in 
Fig.~\ref{fig:Yfcn_modelA_gap10} in which we plot the fluid perturbation 
function $Y$ inside the quark core for the case $\Delta=10$ MeV. 
Note that the value of $n$ corresponds to the number of nodes in the mode 
eigenfunctions.

\begin{figure}
\centering
\includegraphics*[width=8cm]{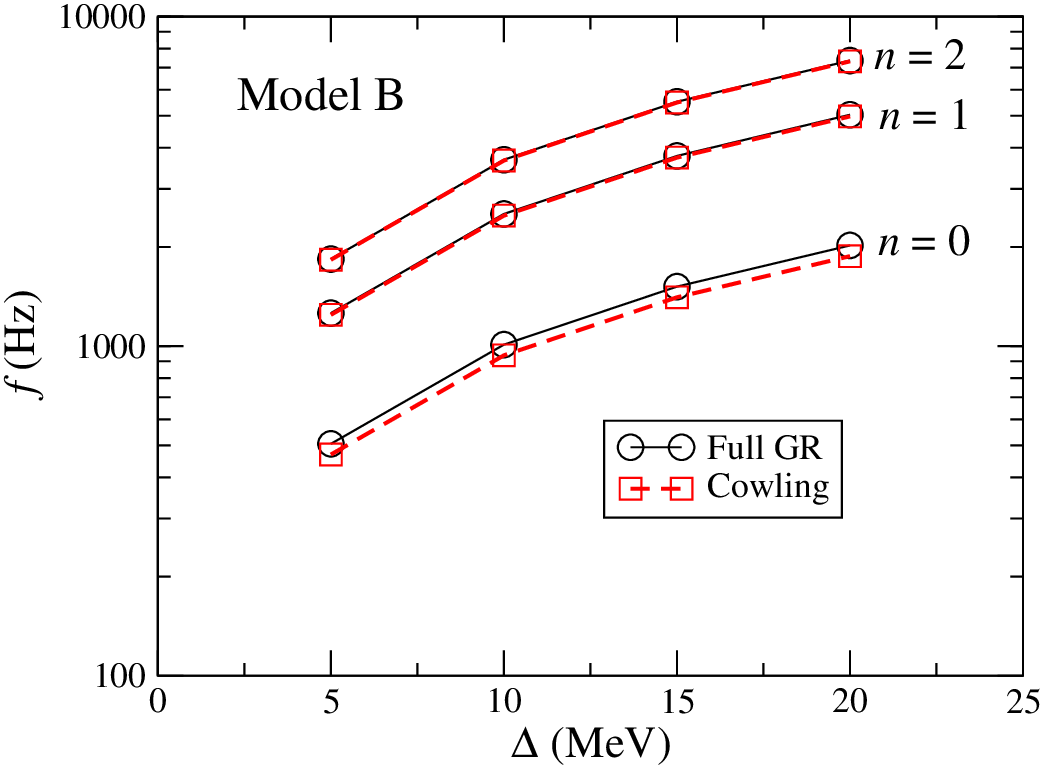}
\caption{Similar to Fig.~\ref{fig:modelA_fvsDelta}, but for the stellar 
model B listed in Table~\ref{tab:models}.}
\label{fig:modelB_fvsDelta}
\end{figure}

In Fig.~\ref{fig:modelA_fvsDelta}, the solid lines are the results obtained 
from the fully general relativistic calculations as discussed in 
Sec.~\ref{sec:perturb_eq}. On the other hand, the dashed lines are obtained by 
using the Cowling approximation where the metric perturbations are neglected 
(see, e.g., \cite{Samuelsson07_p256,Sotani12_p201101} for the relevant 
equations in the Cowling approximation).  
The relative differences between the fully relativistic and Cowling 
results are about 15\% for the $n=0$ fundamental modes. The differences 
are even smaller for the higher harmonics. 
We also note that the relative differences decrease as the size of the quark 
core shrinks. For instance, the fully relativistic and Cowling results for 
the fundamental modes of stellar model C, which has the smallest quark core
among the three models, agree to about 1\%.

Figure~\ref{fig:modelA_fvsDelta} also shows that the mode frequency $f$ depends 
sensitively on $\Delta$. 
In particular, the frequency of the fundamental mode 
($n=0$) increases from 351.8 to 1407.3 Hz as $\Delta$ increases from 5 
to 20 MeV. 
Figure~\ref{fig:modelB_fvsDelta} plots the results of stellar model B for 
comparison. 
In particular, for a given stellar background model, our numerical results 
show that the mode frequency is proportional to $\Delta$. 
This can be understood by noting that the frequency of a shear mode is given 
roughly by $f \sim v_s / 2 \pi R_c$, where the speed of the shear wave is 
$v_s = \left[ \mu / (\rho + P) \right]^{1/2}$ \cite{Samuelsson07_p256}. 
Since the shear modulus $\mu$ is proportional to $\Delta^2$, the mode 
frequency is thus proportional to $\Delta$ as we found numerically.

\begin{figure}
\centering
\includegraphics*[width=8cm]{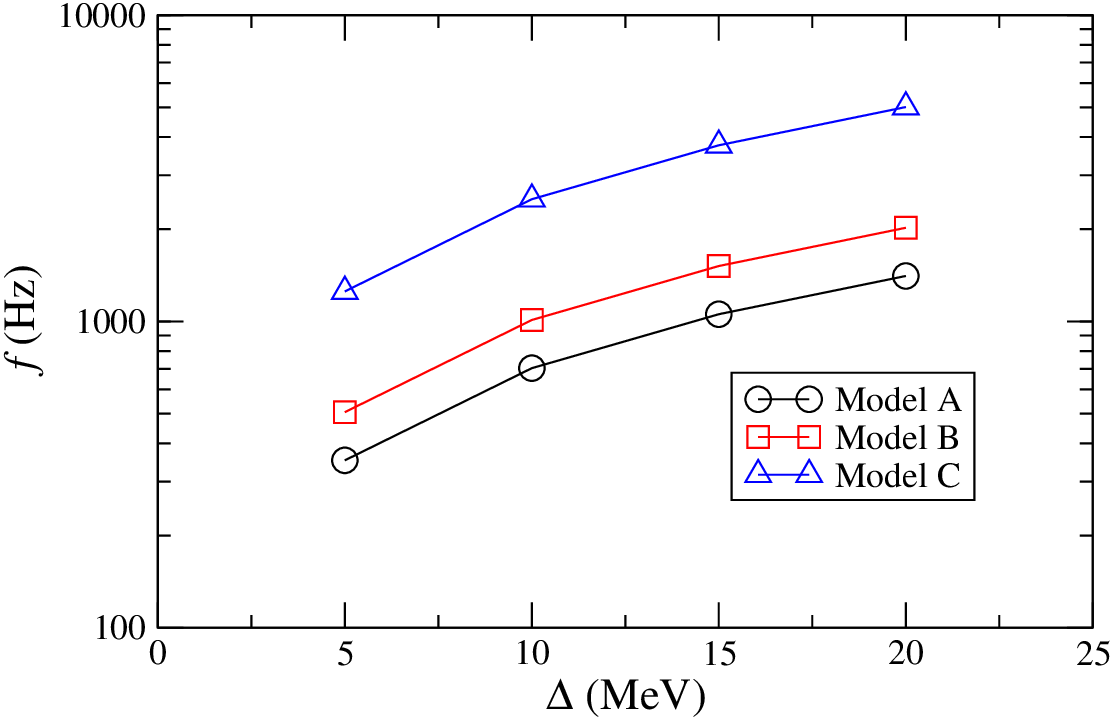}
\caption{ This figure compares the fundamental ($n=0$) mode frequencies 
of the three stellar models listed in Table~\ref{tab:models}. }
\label{fig:1st_mode_Allmodels}
\end{figure}

The dependence of the mode frequency on the size of the quark core (and 
hence the quark-matter EOS parameters) can be seen in 
Fig.~\ref{fig:1st_mode_Allmodels}. 
In the figure, the fundamental modes for the three stellar models are
plotted against $\Delta$. 
As discussed above, models A and C have the largest 
and smallest quark core among the three models, respectively. It is thus 
expected that model A (model C) should have the lowest (highest)
mode frequency for a given value of $\Delta$ as shown in 
Fig.~\ref{fig:1st_mode_Allmodels}, since the speeds of the shear waves for 
the three models do not differ significantly.

It is also interesting to notice that the fundamental-mode frequencies for 
models A and B are in the range from about a few hundred hertz to two
kilohertz. This frequency range is quite different from those of the polar
oscillation modes of traditional neutron stars, such as the $f$ and $p$ modes, 
which have higher frequencies at a few kilohertz or above. 
Hence, these torsional oscillation modes should be easily distinguished from 
those polar oscillation modes. Their relatively low frequencies  
also mean that these modes might already be detectable by the 
second-generation detectors such as Advanced LIGO, since the best 
sensitivities for these detectors are in the range of a few hundred hertz.  
We shall turn to this issue in the next section.

\section{Gravitational-Wave Detection}
\label{sec:GW}

The power carried by any weak gravitational wave of strain 
$h$ is given by (see, e.g., \cite{Sathya09_p2}) 
\begin{equation}
\dot{E} = { c^3 d^2 \over 4 G } {\dot h}^2 , 
\end{equation}
where the dots refer to time derivatives and $d$ is the distance between 
the source and detector. Note that we have restored the constants $G$ and 
$c$. 
Assume that the gravitational-wave signal is an exponentially 
decaying sinusoid with a frequency $f$ and characteristic damping 
time scale $\tau$. The wave strain $h$ can then be expressed as 
\begin{equation}
h = {1 \over 2\pi d f } \left( {4G\over c^3} {E\over \tau} \right)^{1/2} ,
\end{equation}
where we have used ${\dot E} \approx E/ \tau$ and ${\dot h} \approx 
2\pi f h$. Here $E$ is the total energy radiated through the oscillation mode.
Measuring the signal through $N_{\rm c} \approx f \tau$ cycles can boost 
the signal strength by a factor of $\sqrt{N_{\rm c} }$. 
Hence, it is common to define an effective gravitational-wave amplitude 
by $h_{\rm eff} \equiv \sqrt{f\tau} h $: 
\begin{equation}
h_{\rm eff} \approx {1\over 2\pi d} \left( {4G\over c^3}{E\over f} 
\right)^{1/2} . 
\end{equation}
Note that $h_{\rm eff}$ depends only on the energy and frequency of the 
oscillation mode. 

While the torsional oscillation modes are damped very slowly by 
gravitational-wave emission, they could nevertheless be damped much 
faster by internal dissipation. In the Appendix, we provide
an order-of-magnitude estimate for the damping time scale due to internal 
dissipation $\tau_{\rm v}$ and show that 
$f \tau_{\rm v} \sim 10^2 - 10^3$ for the oscillation modes studied by us. 
Assuming $f\tau \gg 1$, the signal-to-noise ratio for the 
gravitational-wave signal can be estimated by \cite{Kokkotas01_p307} 
\begin{equation}
\left( {S \over N} \right) \approx {1\over d f} 
\left[ { G\over 2\pi^2 c^3} 
{ E\over S_h(f)  } \right]^{1/2} ,
\label{eq:SNratio} 
\end{equation}
where $S_h(f)$ is the noise power spectral density of the detector. For the
Advanced LIGO, $S_h(f)$ is given approximately by 
(see Table 1 of \cite{Sathya09_p2})  
\begin{equation}
S_h(f) = S_0 
\left[ x^{-4.14} - 5 x^{-2} + { 111 (1 - x^2 + 0.5 x^4 )\over 1+0.5 x^2}
\right] , 
\end{equation}
where $S_0 = 1.0\times 10^{-49}{\ \rm Hz}^{-1}$ and $x = f/215 {\ \rm Hz}$. 
On the other hand, for a third-generation detector such as the Einstein 
Telescope, $S_h(f)$ is given by (see Table 1 of \cite{Sathya09_p2})
\begin{widetext}
\begin{equation}
S_h(f) = S_0 \left[ x^{-4.1} + 186 x^{-0.69} + 
{ 233 \left( 1 + 31x - 65x^2 + 52x^3 - 42x^4 + 10x^5 + 12 x^6 \right) 
\over 1 + 14x - 37x^2 +19x^3 +27x^4 } \right] , 
\end{equation}    
\end{widetext}
where $S_0=1.5\times 10^{-52} {\ \rm Hz}^{-1}$ and $x = f/200 {\ \rm Hz}$ in 
this case.

\begin{widetext}
\begin{center}

\begin{table}
 \begin{tabular}[t]{|c|c|c|c|c|c|c|}
\hline
 &  &    &    ALIGO   & ALIGO & ET  & ET \\
\hline
Model  & $\Delta$ (MeV)  & $f$ (Hz) & $S/N$ (Crab) & $S/N$ (Vela) 
& $S/N$ (Crab) & $S/N$ (Vela)   \\ 
\hline
A & 5  & 351.8  & $0.3 - 1.0$  & $8.5 - 26.7$ 
  & $4.3 - 13.5$ & $110.3 - 348.8$        \\
  & 10 & 703.7  & $0.05 - 0.2$ & $1.4 - 4.4$ 
  & $1.1 - 3.5$  &  $28.9 - 91.2$  \\ 
  & 20 & 1407.3 & $0.01 - 0.04$ & $0.3-1.0$ 
  & $0.3 - 0.9$  & $7.2 - 22.8$ \\
\hline  
B & 5 & 505.5 & $0.1 - 0.4$ & $3.1 - 9.9$ 
  & $2.1 - 6.8$ & $55.4 - 175.2$ \\
  & 10 & 1010.9 &  $0.02 - 0.08$ & $0.6 - 1.9$ 
  & $0.5-1.7$  &  $14.0 - 44.2$ \\
  & 20 & 2021.9 &  $0.006 - 0.02$  & $0.1 - 0.5$ 
  &  $ 0.1 - 0.4$ & $3.5 - 11.0$ \\
\hline
C & 5 &  1253.5 &  $0.02 - 0.05$ & $0.4 - 1.2$ 
  & $0.4 - 1.1$  &  $9.1 - 28.7$ \\
  & 10 &  2506.9 &  $0.004 - 0.01$ & $0.1 - 0.3$ 
  & $0.09 - 0.3$  & $2.3 - 7.1$  \\
  & 20 & 5013.9 &  $0.0009 - 0.003$  & $0.02-0.07$ 
  & $0.02 - 0.07$  & $0.6 - 1.8$ \\
 \hline
 \end{tabular}
 \caption{ 
Signal-to-noise ratios $S/N$ for the fundamental torsional oscillation modes
of the three stellar models listed in Table~\ref{tab:models}. $S/N$ are 
tabulated for both the Advanced LIGO (ALIGO) and Einstein Telescope (ET). 
$S/N$ (Crab) corresponds to the value for a Crab glitch. The lower (upper) 
limit for each $S/N$ data is obtained by assuming that 10\% (100\%) 
of the glitch energy is channeled to the mode. Similarly, $S/N$ (Vela) 
corresponds the value for a Vela glitch.   } 
 \label{tab:SNratio}
\end{table}

\end{center}
\end{widetext}

\begin{figure}
\centering
\includegraphics*[width=8cm]{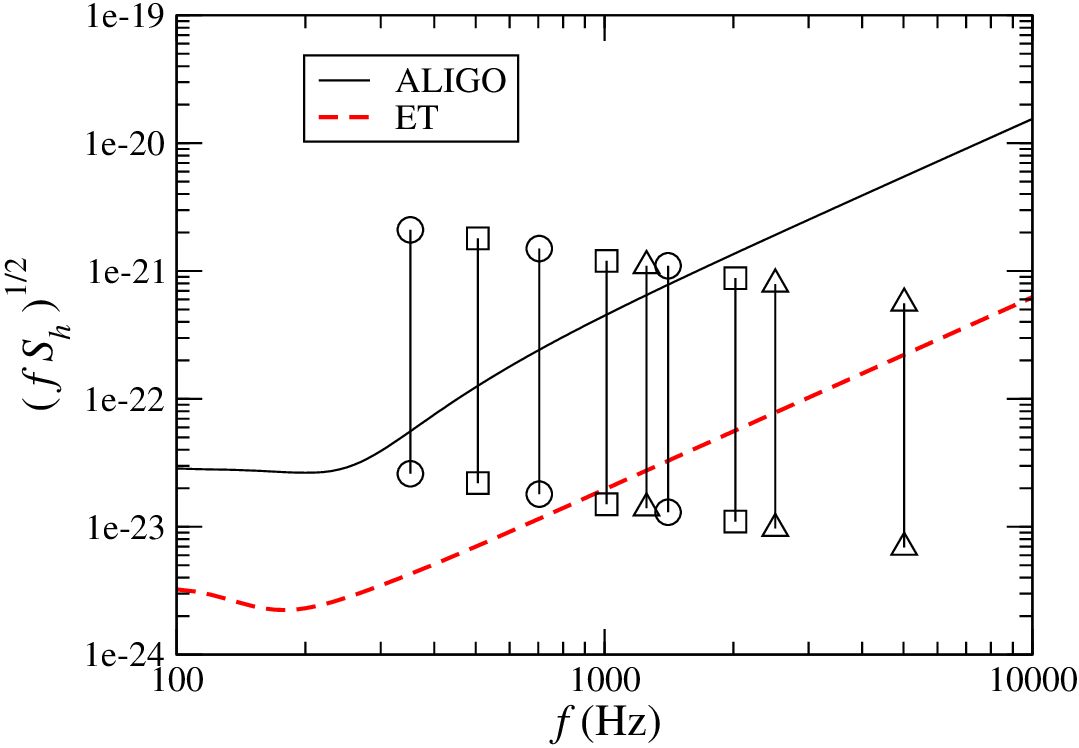}
\caption{ Effective gravitational-wave amplitudes $h_{\rm eff}$ 
(vertical lines) for the modes listed in Table~\ref{tab:SNratio}
are compared to the dimensionless noise 
amplitudes $\sqrt{f S_h}$ for both the Advanced LIGO (ALIGO) and 
Einstein Telescope (ET). For each vertical line, the lower limit is obtained
by assuming that 10\% of the energy of a Crab glitch is channeled to the mode.
The upper limit assumes that 100\% of the energy of a Vela glitch is channeled
to the mode.  }
\label{fig:heff}
\end{figure}

For a given mode frequency and distance to the source, the study of the 
detectability of a gravitational-wave signal now reduces to the problem of 
finding a reasonable estimate for the energy radiated by the oscillation
mode. For the $f$ modes of traditional neutron stars, 
the energy in the mode must 
be at least $\sim 10^{-5} M_{\odot} c^2$ in order for the mode to be 
marginally detectable by Advanced LIGO \cite{Kokkotas01_p307}. 
While this amount of energy in the $f$ mode might be possible for wildly 
pulsating newborn neutron stars formed after supernova explosions, 
this is not the case for mature compact stars with a crystalline quark core 
that we consider in this work. 
A more relevant energy scale for us is the energy associated with a pulsar 
glitch. The released energy can be estimated by \cite{Andersson02_p104002} 
\begin{equation}
E_{\rm glitch} \approx I \Omega \Delta \Omega , 
\end{equation}
where $\Omega$ is the angular velocity of the star and $\Delta\Omega$ is 
the change of $\Omega$ in a glitch event. For simplicity, we shall take 
the moment of inertia $I \sim 10^{45} {\ \rm g \ cm}^2$ for the entire star.
We shall focus on two well-known pulsars, the Crab and Vela pulsars, in the 
following analysis. For the Crab (Vela) pulsar, we have $d=2\ (0.3)$ kpc, 
$\Omega=190.4\ (70.6) {\ \rm rad \ s}^{-1}$, and 
$\Delta\Omega/\Omega = 10^{-8}\ (10^{-6})$. 
The released energies are then estimated to be 
$2\times 10^{-13} M_\odot c^2$ and $3\times 10^{-12} M_\odot c^2$ for the 
Crab and Vela pulsars, respectively. 
Assuming that a torsional oscillation mode excited in a glitch event carries 
a similar amount of energy, we can then estimate the signal-to-noise ratio 
by Eq.~(\ref{eq:SNratio}). 
In Table~\ref{tab:SNratio}, we summarize the signal-to-noise ratios for the 
fundamental $(n=0)$ torsional oscillation modes of the three stellar models 
considered in Sec.~\ref{sec:num_results}. 
We have considered both the Advanced LIGO and Einstein Telescope in the 
calculations. In the table, $S/N$ (Crab) corresponds to the signal-to-noise
ratio for a Crab glitch. Similarly, $S/N$ (Vela) is the value for a 
Vela glitch. 
The lower (upper) limit for each $S/N$ data represents the result obtained by 
assuming that 10\% (100\%) of the glitch energy is channeled to the oscillation
mode.

If we assume that a signal-to-noise ratio must be at least about 10 in 
order for a gravitational-wave signal to be detectable, 
Table~\ref{tab:SNratio} shows that the modes of all three stellar models for 
a Crab glitch are not detectable by Advanced LIGO. 
However, for a Vela glitch, the mode becomes detectable for models A and B
if the gap parameter $\Delta$ is at the lower end of the 
theoretically allowed range near 5 MeV. For stellar model C, which has a 
small quark core, the modes are not detectable by Advanced LIGO for the entire
range of $\Delta$. 
Comparing to Advanced LIGO, the signal-to-noise ratios typically 
increase by 1 order of magnitude for the Einstein Telescope as shown 
in Table~\ref{tab:SNratio}. For this detector, we see that modes for a Crab 
glitch could also be detectable if the quark core is large and 
$\Delta \sim 5$ MeV. For a Vela glitch, modes for stellar models A and B 
should be detectable even for $\Delta\sim 20$ MeV. For model C, however, modes 
excited in a Vela glitch are detectable only if $\Delta \sim 5$ MeV.

One can also see the general trend of our results from Fig.~\ref{fig:heff} in 
which we compare the effective gravitational-wave amplitudes $h_{\rm eff}$ 
(vertical solid lines) for the modes listed in Table~\ref{tab:SNratio} 
with the dimensionless noise amplitudes $\sqrt{f S_h}$ for both detectors.  
For each vertical line in the figure, the lower limit is estimated by 
assuming that 10\% of the energy of a Crab glitch is channeled to the mode.
The upper limit assumes that 100\% of the energy of a Vela glitch is channeled 
to the mode. 
It can be seen that the Einstein Telescope should be able to detect
most of the modes in quite general situations. 
Even more exciting is the possibility that we might already be able to make 
discoveries with Advanced LIGO.

We end this section by discussing briefly how our results would change for 
more massive configurations. 
It should be noted that, for a given EOS model, more massive 
hybrid stars in general have larger quark cores 
(see Fig.~\ref{fig:rho_eosB_diffM}) comparing to those in our canonical 
$1.4 M_\odot$ models. 
As remarked in Sec.~\ref{sec:torsion_modes}, the frequency of a torsional
shear mode is $f \sim v_s / 2 \pi R_c$. 
For a given EOS model and gap parameter, the wave speed 
$v_s = \left[ \mu/ (\rho + P) \right]^{1/2}$ at the surface of 
the quark core is the same for configurations with different masses, since 
the transition density from nuclear matter to quark matter is fixed. 
If we approximate the frequency by evaluating $v_s$ at $R_c$, the frequency of 
a torsional oscillation mode for more massive stars, and hence larger $R_c$,  
should in general be smaller than that of the canonical model with the 
same EOS and gap parameter. 

As seen in Table~\ref{tab:SNratio}, a smaller mode frequency would in general 
increase the signal-to-noise ratio, and hence more massive hybrid stars should 
be more favorable for gravitational-wave detection.  
For instance, the frequency of the fundamental mode of the 
$1.63 M_\odot$ stellar model shown in Fig.~\ref{fig:rho_eosB_diffM} is 
328 Hz for $\Delta = 5$ MeV. This frequency is about a factor of 0.65 smaller 
than the corresponding mode frequency (505.5 Hz) of stellar model B. 
We note that this agrees quite well with our expectation that the mode 
frequency should be smaller by a factor of 0.7 according to the scaling 
relation $f \sim 1/R_c$. 
As a result of a smaller mode frequency, the upper limit of $S/N$ (Vela)
for Advanced LIGO is increased to 32.5 for the $1.63M_\odot$ model, 
which is about three times higher than that of stellar model B. 
Finally, as mentioned in Sec.~\ref{sec:models}, massive hybrid stars 
($M \sim 2 M_\odot$) have been constructed from the NJL model 
\cite{Ippolito08_p023004,Knippel09_p083007}. 
Since the size of the quark core ($R_c \approx 7$ km) and the 
transition density from nuclear matter to quark matter 
($\approx 10^{15}\ {\rm g\ cm}^{-3}$) of the resulting stars are close to 
those of our models (see Fig. 4 of \cite{Knippel09_p083007}), we expect that 
massive stars based on the NJL model would also have similar ranges of 
mode frequency and signal-to-noise ratio as those obtained in our study.

\section{Conclusions}
\label{sec:conclude}

In this paper, we have studied the torsional oscillations of crystalline
color-superconducting hybrid stars in general relativity. The quark matter
in the crystalline core is described by a phenomenological EOS model with 
parameters to account for nonperturbative QCD corrections. 
The low-density envelope of the star is composed of pure nuclear matter and 
the phase transition from quark matter to nuclear matter is implemented by 
using a Maxwell construction. 

We have constructed three canonical $1.4 M_\odot$ stellar models with 
different quark-core sizes and studied their quadrupolar ($l=2$)
torsional oscillation modes in details. 
First we find that the mode frequency depends sensitively on the gap 
parameter $\Delta$ and the size of the quark core. We have also compared 
the results obtained by the fully relativistic calculations with those 
obtained by the Cowling approximation. 
For stellar models with a large quark core ($R_c/R > 0.5$), 
the two results can differ by as much as 15\% for the fundamental modes. 
However, the difference drops down to about 1\% if the star has a 
small quark core ($R_c/R \sim 0.1$). On the other hand, quite independent on 
the quark-core size, the two results in general agree very well for higher 
harmonics. 

Depending on the quark-core size and the value of $\Delta$, the frequency of 
the fundamental torsional oscillation mode can range from a few hundred hertz 
to a few kilohertz in our study. This frequency range is quite different from 
those well-studied oscillation modes of traditional neutron stars, such as 
the $f$ and $p$ modes, with typical frequencies at a few kilohertz. 
Their relatively low frequencies also put some of the torsional oscillation 
modes studied in our models within the best sensitivity region of ground-based 
gravitational-wave detectors.
We have studied the prospect for detecting the gravitational-wave signals 
emitted from these oscillation modes in a pulsar glitch event by assuming
that the energy channeled to the fundamental mode is comparable to the total
energy released in the event. For a Vela glitch, we find that the Einstein 
Telescope should be able to make discovery in quite general situations. 
Furthermore, the signals from a Vela glitch could be detectable by 
Advanced LIGO if the quark core is large ($R_c/R \gtrsim 0.5$) and the gap 
parameter is near the lower limit of the theoretically allowed range at 
$\Delta\sim 5$ MeV. 
Our optimistic results thus suggest the interesting possibility
that we might already be able to probe the nature of crystalline 
color-superconducting quark matter with the second-generation 
gravitational-wave detectors when they come online in the next few years.

Finally, we end this paper with a few remarks. 
(1) As a first investigation of the torsional oscillations of crystalline 
color-superconducting hybrid stars, we simplify our study by using a 
simple phenomenological model to describe the quark matter. In future work, it 
would be interesting to try more realistic models such as the NJL
model of \cite{Ippolito08_p023004,Knippel09_p083007}, which includes the 
three-flavor crystalline phase consistently. 
Nevertheless, we believe that our results would not be 
changed significantly as long as the internal structure of the star 
(e.g., the size of the quark core) is similar to those stellar models
considered in this work. 
Moreover, as we pointed out before, the shear modulus that we 
employ in this work is based on a Ginzburg-Landau expansion to order 
$\Delta^2$, and hence its value only fixes the order of magnitude. Until 
the calculation can be improved, any attempt to calculate the torsional 
oscillation modes of hybrid stars should be regarded as yielding an 
estimate only.
(2) The standard explanation of pulsar glitches is based on the pinning 
and unpinning of superfluid vortices in the inner crust of neutron 
stars \cite{Anderson75_p25,Alpar84_p325,Alpar84_p791}. 
In our analysis of the emitted gravitational-wave signals, we assume that 
glitches are due to the presence of a crystalline quark core inside the star 
so that the fundamental torsional oscillation mode can be excited to a large 
amplitude.  
Future work is needed to study in detail how pulsar glitches may 
be explained by the presence of a crystalline quark core and whether it could
resolve the problem found in \cite{Andersson12_p241103,Chamel13_p011101}, 
where it is shown that superfluid dynamics confined to the inner crust of a 
traditional neutron star is in fact not enough to explain the glitch 
phenomenon. 
(3) We assume that 10\%$-$100\% of the total energy released in a glitch 
event can be channeled to the oscillation mode. 
This is certainly an {\it ad hoc} assumption, but it 
provides us an order-of-magnitude analysis to quantify whether these 
signals might be detectable or not. 
A natural extension of this work is to study how large the amplitude of a 
torsional oscillation mode can be excited in a core-quake scenario
of pulsar glitches \cite{Pines72_p83,Baym76_p829}.


\begin{appendix}

\section{Damping Time Scales}
\label{sec:damping}

In general, there can be different mechanisms acting simultaneously to damp 
an oscillation mode. The damping time scale $\tau$ is controlled mainly by the 
most effective dissipative mechanism in the system. 
As we mentioned at the end of Sec.~\ref{sec:BC}, the torsional oscillations are
damped very slowly by current-quadrupole gravitational-wave emissions. 
The damping time can be estimated by \cite{Schumaker83}
\begin{equation}
\tau_{\rm g} \sim 30 \left( { G M_c \over R_c c^2 }\right)^{-1}
\left( {v_s \over c} \right)^{-5} \omega^{-1} , 
\end{equation} 
where $M_c$ is the mass of the solid core. 
If we take the typical value $\mu \sim 10\ {\rm MeV/fm^3}$ 
[see Eq.~(\ref{eq:mu})] for the shear modulus and evaluate the wave speed 
$v_s$ at the surface of the quark core $R_c$, we would obtain
$f \tau_{\rm g} \sim 10^6$ (with $f$ being the mode frequency)
and thus gravitational radiation is not an 
effective damping mechanism for the torsional oscillations. The fact that 
$f \tau_{\rm g} \gg 1$ is the reason why the deep minimum in 
Fig.~\ref{fig:modelA_gap10_3rdmode} has such a narrow width.

The oscillation modes are also damped by internal microphysics processes. 
Dissipative mechanisms in color-superconducting quark matter have been 
studied mainly for the CFL phase of quark matter 
(e.g., \cite{Manuel05_p076,Alford07_p055209,Mannarelli08_p241101,
Mannarelli10_p043002,Andersson10_p023007}). 
At very low temperatures, it is expected that dissipative processes are 
dominated by phonon-phonon scattering. In particular, the coefficient for
the phonon shear viscosity in the CFL phase is \cite{Manuel05_p076}
\begin{equation}
\eta  \approx 2.5 \times 10^{27} 
\left( { \mu_q \over 400\ {\rm MeV} } \right)^8
\left( { T \over 10^9\ {\rm K} } \right)^{-5}\ {\rm g\ cm^{-1} s^{-1}} . 
\label{eq:cfl_shear}
\end{equation}
On the other hand, dissipative mechanisms in the crystalline phase of quark 
matter have not been studied (as far as we are aware). 
We shall proceed by applying the formalism developed in
\cite{Landau} where it is shown that the viscosity of a solid 
can be treated formally as that of a fluid. 
The rate of energy dissipation is given by \cite{Landau}
\begin{equation}
{d E \over dt} = - 2 \int f_{\rm dis} d V ,
\label{eq:dEdt}  
\end{equation}
where the dissipation function $f_{\rm dis}$ is defined by 
\begin{equation}
f_{\rm dis} = \eta \Sigma^{ij} \Sigma_{ij} + {1\over 2} \zeta \Theta^2 . 
\label{eq:f_dis}
\end{equation} 
The shear tensor $\Sigma_{ij}$ is 
\begin{equation} 
\Sigma_{ij} = {1\over 2} \left( \nabla_i v_j + \nabla_j v_i \right) 
- {1\over 3} g_{ij} \Theta , 
\end{equation} 
where $v_i$ is the velocity and the expansion $\Theta = \nabla_i v^i$. 
Note that the coefficients $\eta$ and $\zeta$ in Eq.~(\ref{eq:f_dis}) 
corresponds to the shear and bulk viscosities, respectively. Given the rate
of energy dissipation (\ref{eq:dEdt}), we can estimate the damping time scale
of an oscillation mode due to internal viscosities by 
$\tau_{\rm v} = 2 E / \left| {dE \over dt} \right|$, 
where the energy of the mode is $E =  \int \rho v^2 dV / 2$ 
(see, e.g., \cite{Andersson10_p023007}). 
Since the amplitude of the fundamental torsional 
oscillation mode reaches a maximum at the surface of the quark core 
(see Fig.~\ref{fig:Yfcn_modelA_gap10}), we shall approximate 
$E$ and $dE/dt$ by evaluating their integrands at $r=R_c$. 
Restricting to the torsional oscillations, where the velocity field is 
divergence-free, we obtain the following estimate for the damping time scale:
\begin{equation} 
\tau_{\rm v} \sim { \rho(R_c) R_c^2 \over \eta } , 
\end{equation}
where $\rho(R_c)$ is the density at the surface of the solid core and we have 
also neglected numerical factors of order unity. 
The value of the coefficient $\eta$ has not been computed for the crystalline 
quark matter. As a rough estimate, we take the typical value 
$\eta \approx 10^{27}\ {\rm g\ cm^{-1} s^{-1}}$ as suggested in
Eq.~(\ref{eq:cfl_shear}) for the CFL phase at $T=10^9$ K. 
Using the typical values $\rho(R_c) \approx 10^{15}\ {\rm g\ cm^{-1}}$ 
and $R_c \approx 5$ km for our canonical models, we obtain the damping 
time scale due to internal dissipation $\tau_{\rm v} \sim 0.3$ s. 
Comparing to gravitational-wave damping, our analysis suggests that the 
torsional oscillation modes would be damped much faster by internal 
dissipation. However, for the range of mode frequency we studied 
(see Table~\ref{tab:SNratio}), we obtain $f \tau_{\rm v} \sim 10^2 - 10^3$ and 
hence $f \tau \gg 1$ is still a reasonable assumption.

Our analysis is a crude approximation based on the assumption 
that phonon-phonon scattering in the crystalline phase would not differ 
significantly from that in the CFL phase. 
More detailed work in this direction is needed to improve our estimation. 
In fact, as pointed out in \cite{Manuel05_p076}, the hydrodynamic treatment of 
the phonons used in the derivation of Eq.~(\ref{eq:cfl_shear}) is not valid 
anymore in the low-temperature regime, where the mean-free path of phonons 
would be much larger than the stellar radius. The CFL quark matter would then
become a perfect superfluid. If we use the effective shear viscosity suggested 
in \cite{Andersson10_p023007}, 
$\eta_{\rm eff} \approx 5 \times 10^{17} \left( T/10^9\ {\rm K} \right)^4
\ {\rm g\ cm^{-1}s^{-1} }$, the value of which is obtained by requiring 
that the mean-free path of phonons is limited by the stellar radius, then 
the damping time scale at the same temperature $10^9$ K would be increased 
by a factor of $10^9$.

\end{appendix}


\bibliographystyle{prsty}

\end{document}